\documentclass[prl,twocolumn,floatfix,superscriptaddress,byrevtex]{revtex4}
\usepackage[english]{babel}
\usepackage{bm,amsmath,amssymb}
\usepackage{dcolumn}
\usepackage{graphicx}
\begin{document}
\title{Dynamical slowdown of polymers in laminar and random flows}
\author{A. Celani}
\author{A. Puliafito}
\affiliation{CNRS--INLN, 1361 Route des Lucioles, 06560 Valbonne, France}
\author{D. Vincenzi}
\affiliation{Max-Planck-Institut f\"ur Dynamik und Selbstorganisation,
Bunsenstra{\ss}e 10, 37073 G\"ottingen, Germany}
\affiliation{Sibley School of Mechanical \& Aerospace Engineering and
LASSP,
Cornell University, Ithaca, NY 14853}
\begin{abstract}
The influence of an external flow on the relaxation dynamics of a
single polymer is investigated theoretically and numerically.
We show that a pronounced dynamical slowdown occurs in the vicinity of 
the coil--stretch transition, especially when the dependence on 
polymer conformation of the drag is accounted for.
For the elongational flow, relaxation times are
exceedingly larger than the Zimm relaxation time, 
resulting in the observation of conformation hysteresis.  
For random smooth flows hysteresis is not present. Yet, 
relaxation dynamics is significantly
slowed down because of the large variety of accessible polymer 
configurations. The implications of these results for the modeling  
of dilute polymer solutions in turbulent flows are addressed.
\begin{center}
{\fontfamily{ppl}\selectfont{\itshape Phys. Rev. Lett.}
\textbf{97}, 118301 (2006)
\\ \texttt{http://link.aps.org/abstract/PRL/v97/e118301}
}
\end{center}
\end{abstract}
\pacs{}
\maketitle
The dynamics of an isolated polymer in a flow field
forms the basis of
constitutive models for dilute polymer solutions~\cite{chu,larson,shaqfeh}. 
The modeling of drag-reducing flows, for instance,
requires an appropriate description
of single polymer deformation in turbulent velocity fields~\cite{95GB}.
In the last decade, major advances in fluorescence
microscopy offered the possibility
of tracking isolated polymers both in laminar~\cite{shaqfeh} 
and random flows~\cite{05GCS}.
The dynamics of a polymer in thermal equilibrium with the
surrounding solvent is commonly described in terms of 
normal modes and relaxation times associated with them.
The analytical form of the relaxation spectrum was first obtained 
by Rouse under the assumption that the polymer could be described
as a series of beads connected by Hookean springs~\cite{53R}.
The Rouse model was subsequently improved by Zimm to include hydrodynamical 
interactions between segments of the polymer~\cite{zimm}. In Zimm's 
formulation,
the equations of motion are decoupled into a normal mode structure by
preaveraging the distances between the beads over the distribution of polymer
configurations.
The relaxation time associated with the fundamental mode, $\tau$, determines
the typical time that it takes for a deformed polymer to recover the 
equilibrium configuration in a solvent.
The normal mode theory was confirmed 
by the analysis of the oscillatory motion of a DNA molecule
immersed in a solvent and held in a partially extended state 
by means of optical tweezers~\cite{97QBC}. An alternative approach 
to examine polymer relaxation consists in stretching a
tethered DNA molecule in a flow and measuring 
its relaxation after cessation of the flow~\cite{94PQSC,96MCVCC}.
The theoretical predictions for these experimental conditions
are provided by the scaling theory~\cite{95MB} 
and the static dynamics formalism~\cite{02RZ}.

The aforementioned studies all consider
the internal dynamics of a polymer floating in a solvent under the influence
of thermal noise only ---
the interaction of the polymer with an external flow is not taken into account.
One of the aspects highlighted by experiments is that polymer dynamics
in a moving fluid is strongly influenced by the carrier velocity field.
The coil--stretch transition  is the most noticeable example:
as  the strain rate
exceeds a threshold value, the polymer undergoes a transition 
from the coiled, equilibrium configuration to an almost fully extended 
one~\cite{74DG}.
Therefore, when a polymer is freely transported by a non-uniform flow,
we expect that the time scales describing its dynamics 
may be significantly different from the Zimm time~$\tau$.
Simple models for polymer stretching indeed
suggest deviations from Zimm's theory near the 
coil--stretch transition~\cite{05CMV,05MV,05VB}. 
Discrepancies in the definition of the 
correct relaxation time are also encountered in drop formation 
experiments~\cite{TKCW}. 

In this Letter we investigate polymer relaxation dynamics
both in elongational and random smooth flows. Our analysis
brings to evidence an important slowdown
of dynamics with respect to
the Zimm timescales, in the vicinity of the coil--stretch transition. 
For the elongational flow, this is related 
to conformation hysteresis~\cite{03Schetal,04SSC,05HL}. 
For random flows, we show that
hysteresis is not present. Nonetheless, the amplification 
of the relaxation time persists, albeit to a lesser extent, 
due to the large heterogeneity of polymer
configurations. In both cases, the dependence of the drag force on
the polymer configuration plays a prominent role. This suggests
the necessity of improving current models 
of polymer solutions in turbulent flows to account for such effect.

The dumbbell approximation is the basis of the 
most common models of single polymer dynamics 
and  viscoelastic models of dilute polymer solutions \cite{Bird}.
Its validity relies on the fact that the slowest deformational mode
of the polymer is the most influential in producing 
viscoelasticity~\cite{95GB}.
However, when attention is directed to non-equilibrium dynamics it
is often too crude to assume
that~$\tau$ is independent of the conformation of the molecule~\cite{99HQ}.
Therefore, following de Gennes~\cite{74DG},
Hinch~\cite{74H}, and Tanner's~\cite{75T} approach, we consider a model where
the polymer is described as two beads connected by an elastic spring
and the probability density function of the end-to-end vector, $\Psi(\bm{R},t)$, 
satisfies the diffusion equation:
\begin{equation} \label{eq:dumbbell}
\frac{\partial\Psi}{\partial t}=
-\frac{\partial}{\partial \bm R}\cdot\bigg[\bm\kappa(t)\cdot\bm R\,\Psi-\frac{f(R)\bm R}{2\tau\nu(R)}\,\Psi
-\frac{R_0^2}{2\tau \nu(R)}\,\frac{\partial\Psi}{\partial\bm R}
\bigg]
\end{equation}
where $\kappa_{ij}(t)=\partial_jv_i(t)$~is the velocity gradient, $R_0$ is the mean extension at 
thermal equilibrium and~$R=|\bm{R}|$.
The function~$f(R)$ defines the entropic force restoring stretched molecules
into the coiled configuration. Synthetic polymers 
are properly described
by the Warner law, $f(R)=1/(1-R^2/L^2)$, where~$L$ is the contour length of
the polymer; biological macromolecules
are better characterized by the Marko--Siggia law, 
$f(R)=2/3-L/(6R)+L/[6R(1-R/L)^2]$~\cite{larson}.
The flow strength relative to the polymer tendency to recoil is
expressed by the Weissenberg number~\textit{Wi},
defined as the product of the Zimm time~$\tau$ by a characteristic
extension rate of the flow.
The function~$\nu(R)$ encodes for the dependence on the polymer 
conformation of the drag exerted by the fluid: 
a spherical coil offers a smaller resistance with respect to a 
long rod-like configuration.
We utilize the expression~$\nu(R)=1+(\zeta_{s}/\zeta_{c}-1)R/L$
that interpolates linearly between these extremes (see Refs.~\cite{97L,Doyle}). Here, $\zeta_{c}= 3\sqrt{6\pi^3}R_0\eta/8$ and 
$\zeta_{s}= 2\pi L \eta/\ln(L/\ell)$ are the friction coefficients
for the coiled and the stretched configuration, respectively,
$\eta$ is the solvent viscosity and $\ell$ is the diameter of the molecule. 
Strictly speaking, the above form of~$\nu(R)$ was deduced from 
experiments and microscopic simulations of laminar flows.
To our knowledge, measures of~$\nu(R)$ in random flows
are not available yet, and the study of its functional dependence lies beyond
the scope of the present work.  
Assuming that the polymer is aligned with the local 
stretching direction of the flow for the most part of its evolution,
we shall  use a linear~$\nu(R)$ for a random flow as well. 

To define the relaxation time in the presence of an arbitrary
external flow, we consider the probability density function of the 
rescaled extension, $P(r,t)$ with $r=R/L$,
and identify the dynamical relaxation time, $t_{\mathrm{rel}}$,
as the characteristic time needed for~$P(r,t)$ to attain its stationary 
form~$P_{\mathrm{st}}(r)$. In the cases examined in this Letter, 
as we shall see,
the probability density function of~$r$ satisfies the 
Fokker--Planck equation
\begin{equation}\label{eq:fp}
\partial_{t'} P=-\partial_r(D_1(r)P)+
\partial_rD_2(r)\partial_r P, 
\end{equation}
where the form 
of~$D_1(r)$ and~$D_2(r)$ depends on the flow and $t'=t/\tau$. 
The stationary solution to Eq.~\eqref{eq:fp}
takes the potential form 
$P_{\mathrm{st}}(r)=N\exp{[-E(r)/K_BT]}$, where 
$N$ is the normalization constant and
$E(r)=-K_BT\int^r D_1(\rho)/D_2(\rho)\,d\rho$.
The finite-time solution admits the expansion
\begin{equation}\label{eq:tdfp}
P(r,t')=P_{\mathrm{st}}(r)
+\sum_{n=1}^{\infty}a_n \mathit{p}_n(r)\,  e^{-t'/\sigma_n},
\end{equation}
where the coefficients~$a_n$ are fixed by~$P(r,0)$,
$p_n(r)$ are the eigenfunctions of the Fokker--Planck operator,
and~$\sigma_n$ are the reciprocals of its (strictly positive) eigenvalues,
arranged in descending order ($\sigma_n > \sigma_{n+1}$). 
The relaxation time is thus defined as  $t_{\mathrm{rel}}\equiv\sigma_1 \tau$.
\begin{figure}
\includegraphics[width=.45\columnwidth]{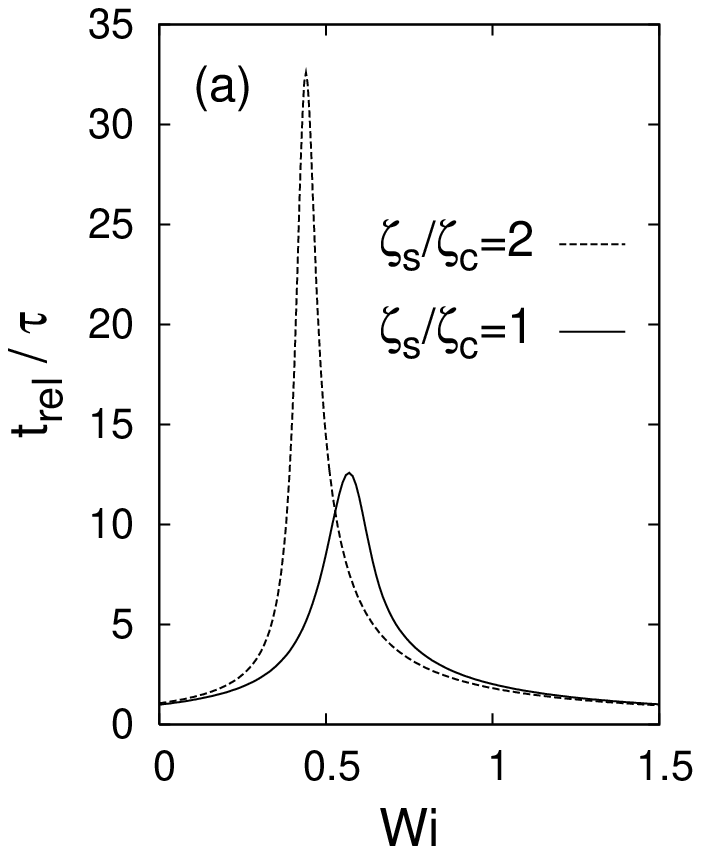}\hfill
\includegraphics[width=.45\columnwidth]{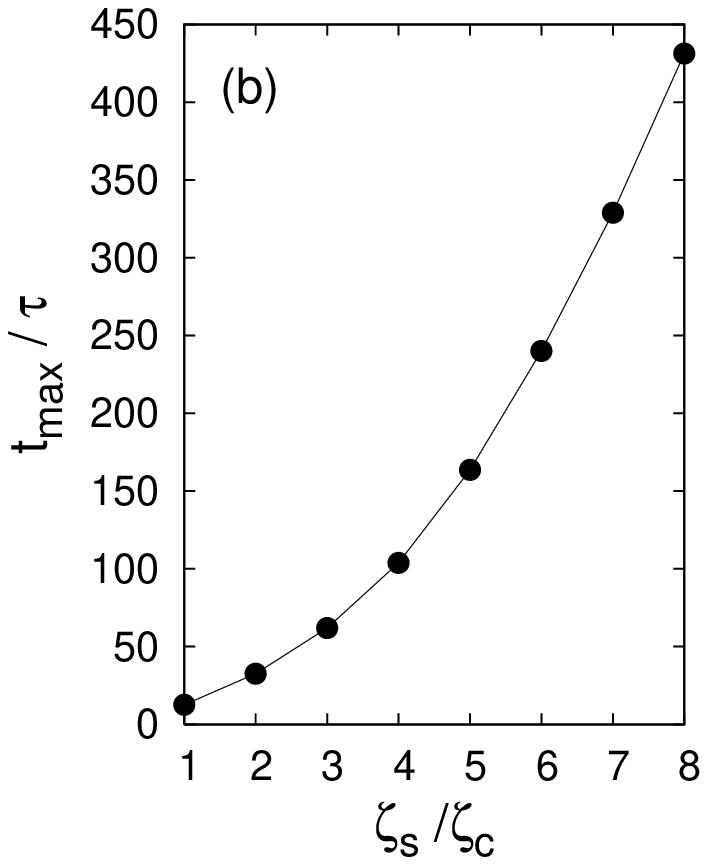}
\caption{Elongational flow: (a)
rescaled relaxation time vs.~$\textit{Wi}$ for $b=400$. 
The entropic force is given by the Warner law.
For small~$\textit{Wi}$, $t_{\mathrm{rel}}/\tau$
grows as $1/(1-2\textit{Wi})$, as can be seen by replacing~$\hat{f}(r)$
with~1 in Eq.~\eqref{eq:fp}.
For~$\textit{Wi}\gg\textit{Wi}_{\text{crit}}$, 
the time needed to reach the asymptotic regime is set by the time
scale of the flow, $\gamma^{-1}$, and~$t_{\mathrm{rel}}/\tau$ decreases 
as~$\textit{Wi}^{-1}$.
(b)~Rescaled maximum relaxation time~$t_{\text{max}}/\tau$
vs.~$\zeta_{s}/\zeta_{c}$ ($b=400$).}
\label{fig:relax}
\end{figure}

\begin{figure}[t]
\includegraphics[width=0.49\columnwidth]{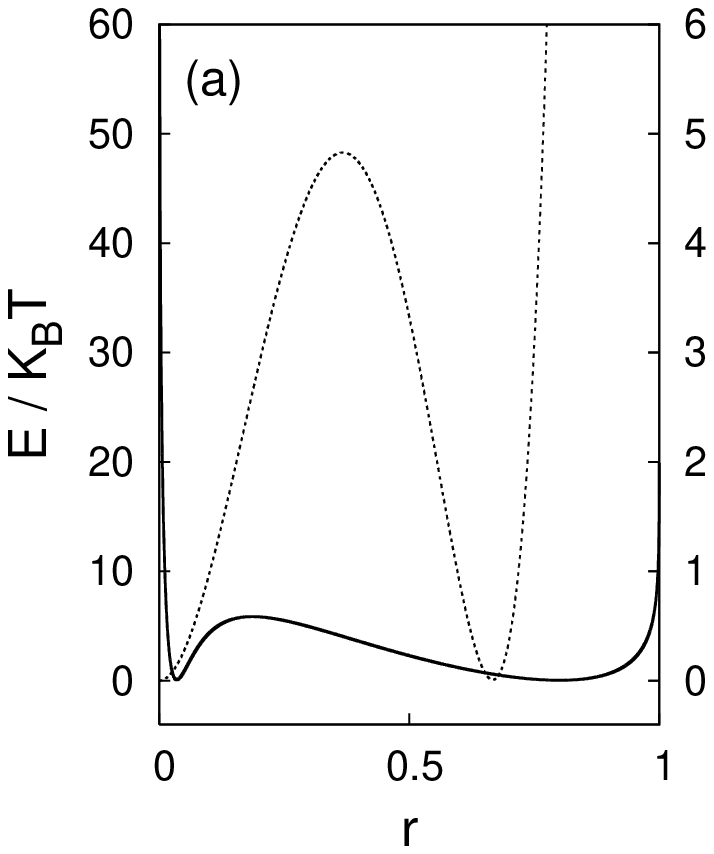}\hfill
\includegraphics[width=0.465\columnwidth]{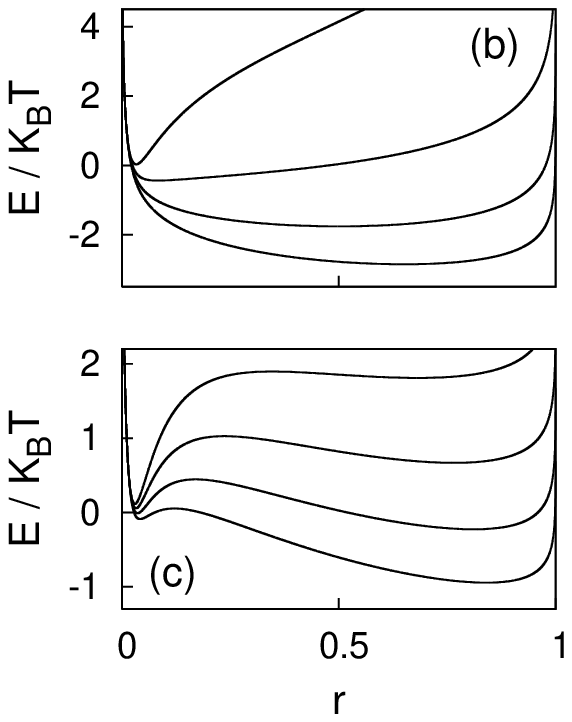}
\caption{(a) Effective energy at the coil--stretch
transition for a polyacrylamide (PAM) molecule 
($b=3953$, $\zeta_s/\zeta_c=6.87$) in the elongational flow (dashed line) 
and the 3D Batchelor--Kraichnan flow (solid line).
The left vertical axis refers to the dashed line; the right vertical axis 
refers to the solid line; (b) effective energy 
in the random flow
for a PAM molecule with constant drag ($\zeta_s$=$\zeta_c$)
(from top to bottom~$\textit{Wi}=0.4,0.7,1.0,1.3$); (c) the same as (b),
but with~$\zeta_s=6.87\zeta_c$
(from top to bottom~$\textit{Wi}=0.28,0.33,0.38,0.43$).
When~$\zeta_s>\zeta_c$
the transition occurs in a much narrower range of~\textit{Wi}.
}
\label{fig:potential}
\end{figure}

As a first example, we examine the steady planar
elongational flow~$\bm v=\gamma(x,-y)$.
By assuming that the polymer extension in the $x$-direction is much greater
than in the $y$-direction, it is easy to derive from Eq.~\eqref{eq:dumbbell} 
an equation of the form~\eqref{eq:fp}
with $\displaystyle D_1(r)=\mathit{Wi}\,r-\hat{f}(r)r/[2
\hat{\nu}(r)]$, $\displaystyle D_2(r)=[2b\hat{\nu}(r)]^{-1}$, 
$\hat{f}(r)=f(rL)$, $\hat{\nu}(r)=\nu(rL)$, $b=L^2/R_0^2$,
and~$\textit{Wi}=\gamma\tau$~\cite{note}.
For~$\zeta_{s}=\zeta_{c}$, $t_{\mathrm{rel}}$ can be computed 
from~\eqref{eq:tdfp} by solving a central two-point 
connection problem for a generalized spheroidal wave 
equation~\cite{05VB}. In the general case, $\zeta_{s}>\zeta_{c}$,
we resorted to a numerical
computation based on the variation--iteration method of quantum 
mechanics~\cite{MF}.
In the vicinity of the coil--stretch transition ($\mathit{Wi}=1/2$)
$t_{\mathrm{rel}}$ shows a sharp peak (Fig.~\ref{fig:relax}).
In this range of~\textit{Wi}
there is a critical competition between the entropic force and the
velocity gradient 
that makes the convergence time to the steady state extremely 
long. This effect is strongly enhanced by the conformation-dependent drag;
the peak~$t_{\text{max}}$ indeed increases 
with~$\zeta_{s}/\zeta_{c}$~(Fig.~\ref{fig:relax}). 
Those extremely long relaxation times are intimately connected with
the observation of finite-time conformation 
hysteresis~\cite{03Schetal,04SSC,05HL}.
For large enough $\zeta_{s}/\zeta_{c}$, there is a
narrow range of~\textit{Wi} around the critical value for the 
coil--stretch transition where~$E(r)$ has a double well
structure~\cite{03Schetal,04SSC,05HL}.
The barrier height separating the coiled and the stretched state
is much greater than the thermal energy~$K_BT$,
and therefore the polymer remains trapped in its initial
configuration for an exceptionally long time (Fig.~\ref{fig:potential}).

\begin{figure}[b]
\includegraphics[width=0.45\columnwidth]{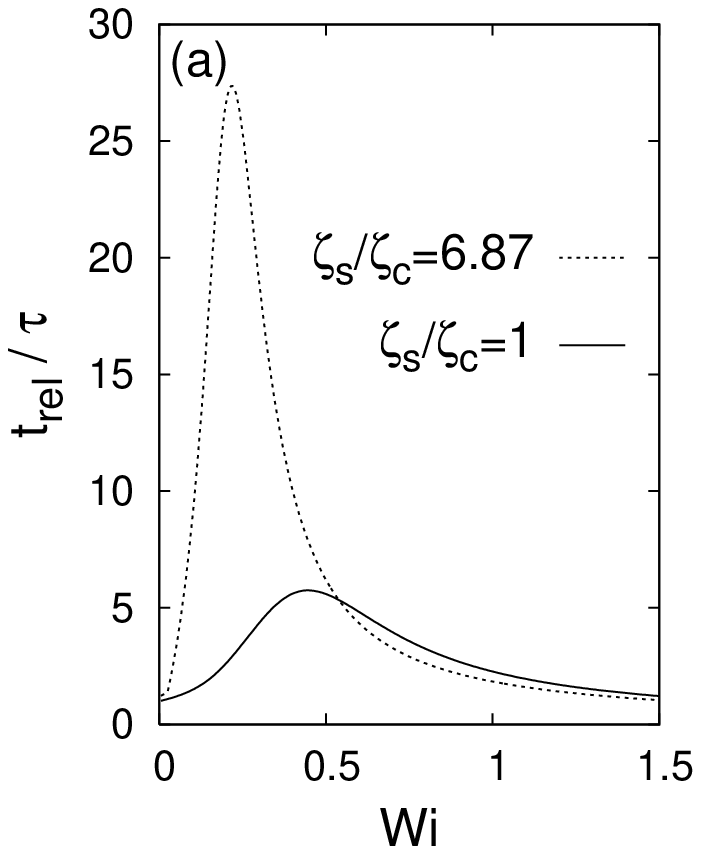}\hfill
\includegraphics[width=0.435\columnwidth]{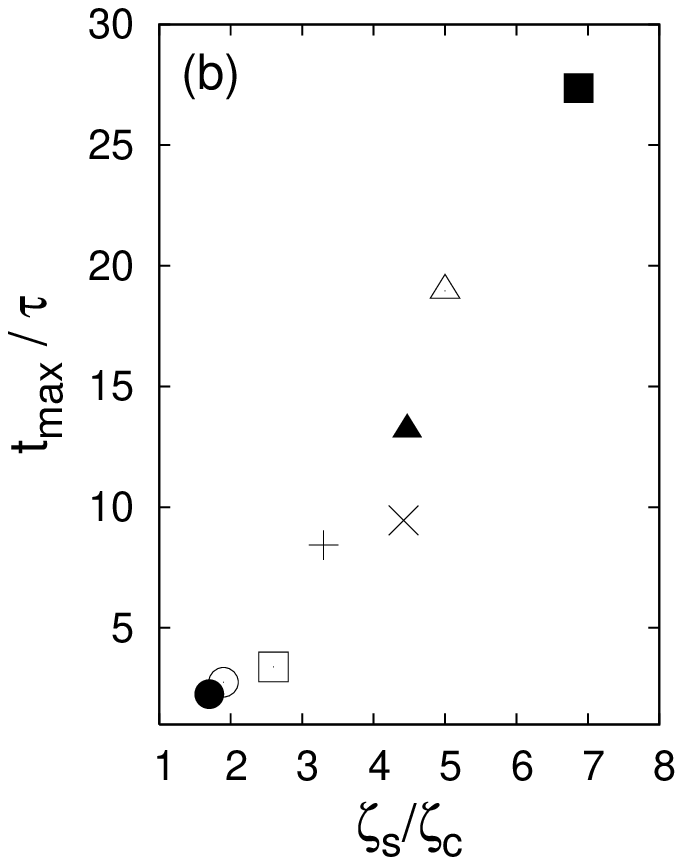}
\caption{3D Batchelor--Kraichnan flow: 
(a) $t_{\mathrm{rel}}/\tau$ vs.~$\mathit{Wi}$
for a PAM molecule ($b=3953$); (b)~$t_{\text{max}}/\tau$ for the following polymers:
DNA ($\bullet$, $b=191.5$; $\circ$, $b=260$; $\square$,
$b=565$; $+$, $b=2250$), 
polystyrene ($\times$, $b=673$), 
polyethyleneoxide (PEO) ($\blacktriangle$, $b=1666$), 
Escherichia Coli DNA ($\vartriangle$, $b=9250$), 
PAM ($\blacksquare$). Measures of~$b$ and~$\zeta_{s}/\zeta_{c}$ 
can be found in~\cite{97L,03Schetal,04SSC,05HL}.
Synthetic polymers are modeled
by the Warner law, whereas biological molecules are described by the 
Marko--Siggia law. Relaxation times were computed by means of 
the variation--iteration method \cite{MF}. 
For~$\zeta_{s}=\zeta_{c}$ they
can be obtained by solving
an eigenvalue problem for a  Heun equation~\cite{05MV}.}
\label{fig:tdiz}
\end{figure}

The discovery of elastic turbulence has recently
allowed the examination of single
polymer dynamics in a random smooth flow generated by viscoelastic 
instabilities~\cite{00GS,05GCS}. 
The velocity gradient fluctuates along 
fluid trajectories; the average local deformation rates define the three
Lyapunov exponents of the flow.
Experimental observations have been accompanied by theoretical and
numerical studies based on the dumbbell model~
(see, e.g., Refs.~\cite{Chertkov00,05CMV,05MV}).
To analytically investigate polymer relaxation dynamics in 
random flows, we initially
assume that the velocity field obeys the Batchelor--Kraichnan 
statistics~\cite{68K}.
The velocity gradient is then a statistically isotropic and 
parity invariant Gaussian process with zero mean and
correlation: $\langle \kappa_{ij}(t)\kappa_{kl}(s)\rangle=
2\lambda\delta(t-s)[(d+1)\delta_{ik}\delta_{jl}-\delta_{ij}\delta_{kl}
-\delta_{il}\delta_{jk}]/[d(d-1)]$, where~$d$ is the dimension of the
flow and~$\lambda$ denotes the largest Lyapunov exponent.
In this context, we indicate by~$P(r,t')$ 
the probability density function of the extension 
both with respect to thermal noise and the realizations 
of the velocity field.
For the elongational flow Eq.~\eqref{eq:fp} was obtained under the
uniaxial approximation. On the contrary,
for the isotropic random flow Eq.~\eqref{eq:fp}
holds exactly and can be obtained by a Gaussian integration by parts followed
by integration over angular variables. 
The drift and diffusion coefficients take the form
$\displaystyle D_1(r)=(d-1)/d\,\textit{Wi}
\;r-\hat{f}(r)r/
[2\hat{\nu}(r)]+(d-1)/[2b\hat{\nu}(r)r]$ and 
$\displaystyle D_2(r)=\textit{Wi}\;
r^2/d+[2b\hat{\nu}(r)]^{-1}$ with~$\mathit{Wi}=\lambda\tau$.
The stationary distribution admits once more a potential form.
For~$d=3$ and~$\zeta_{s}>\zeta_{c}$, the potential
displays a very wide well, 
the effect of the 
conformation-dependent drag being to increase the probability of large
extensions. 
There is no evidence of pronounced double wells  (Fig.~\ref{fig:potential}). 
Near the coil--stretch transition, the effective 
barrier heights separating the coiled and 
stretched states
are indeed at most comparable to thermal energy.
For realistic~$\zeta_{s}/\zeta_{c}$
no conformation  hysteresis is therefore expected to be observed in 
random flows. The behavior of~$t_{\mathrm{rel}}$ 
vs.~$\mathit{Wi}$ is however analogous to the one encountered
in the elongational 
flow: $t_{\mathrm{rel}}/\tau$
increases as $[1-\mathit{Wi}(d+2)/d]^{-1}$ at small~\textit{Wi} and
decreases as~$\mathit{Wi}^{-1}$ at large~\textit{Wi}.
A peak near the transition is present, 
that becomes more and more pronounced
with increasing~$\zeta_{s}/\zeta_{c}$,
attaining values as large as about thirty times~$\tau$
(Fig.~\ref{fig:tdiz}). 
The reason for this
behavior is the breadth of~$P_{\mathrm{st}}(r)$ and
the consequent large heterogeneity of accessible polymer configurations.

\begin{figure}[t]
\includegraphics[width=0.47\columnwidth]{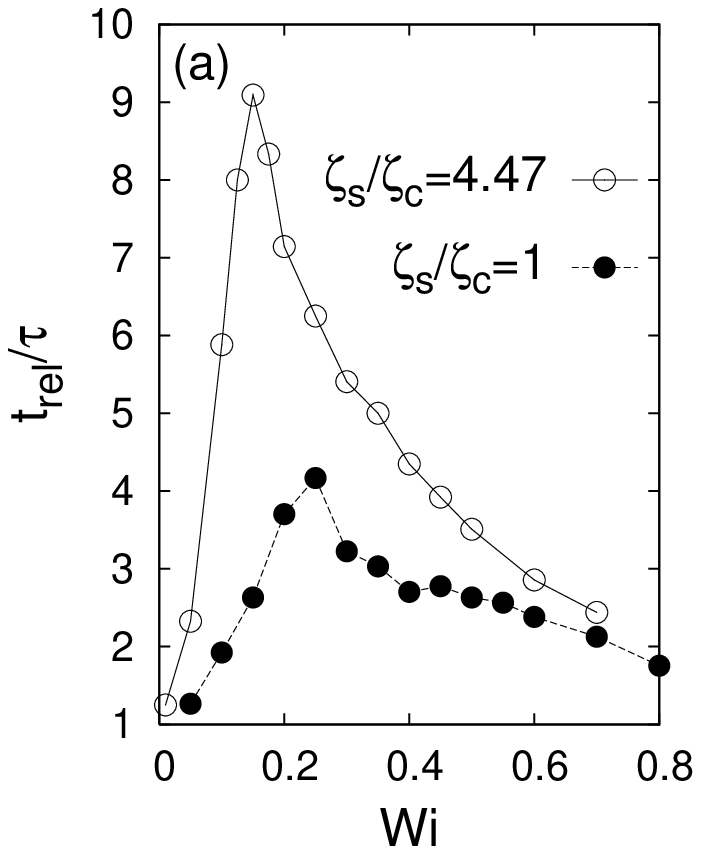}\hfill
\includegraphics[width=0.47\columnwidth]{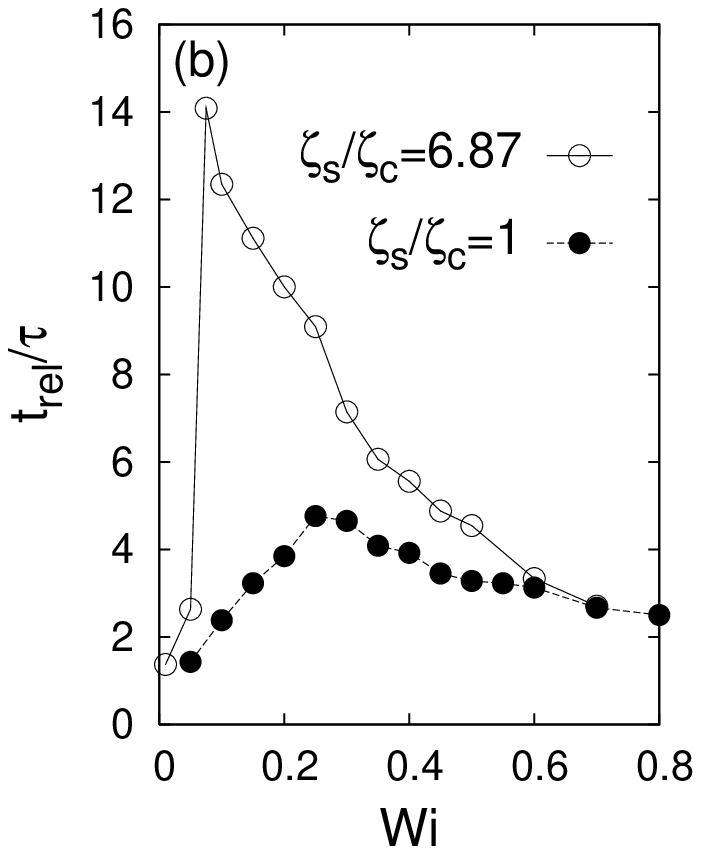}
\caption{Brunk--Koch--Lion flow:
$t_{\mathrm{rel}}/\tau$ vs.~$\textit{Wi}=\lambda\tau$ 
from Brownian Dynamics simulations; \label{fig:numerics} (a)~PEO; 
(b)~PAM.}
\end{figure}

To corroborate the results obtained in the context of
the short-correlated flow, we performed Brownian Dynamics 
simulations of dumbbell molecules~\cite{Ito} in the random flow 
introduced by Brunk et al.~\cite{brunk}. This model reproduces
the small-scale structure of a turbulent flow by means of a 
statistically isotropic Gaussian velocity gradient. 
The autocorrelation times of components of the strain and rotation tensors
are set to be multiple of the Kolmogorov time~$\tau_\eta$
by comparison with direct numerical simulations of 3D isotropic turbulence
($\tau_S=2.3\tau_\eta$, $\tau_R=7.2\tau_\eta$).
The Lyapunov exponent of this flow is 
$\lambda\simeq 10\tau_{\eta}^{-1}$. We computed~$t_{\mathrm{rel}}$ as
the time of convergence of the moments $\langle r^n(t)\rangle$ to
their stationary value~$\langle r^n\rangle_\mathrm{st}$:
$t_{\mathrm{rel}}^{-1}=-\lim_{t\to\infty}\ln{[\langle r^n(t)\rangle
-\langle r^n\rangle_{\mathrm{st}}]}/t$, where the averages were taken
over an ensemble of realizations of the flow,
all with the same initial extension~$r(0)$.
The numerical difficulty arising from the singularity of the entropic 
force at~$R=L$ has been overcome by exploiting the algorithm introduced 
in~\cite{05CPT}. 
The results shown in Fig.~\ref{fig:numerics} confirm 
the scenario depicted in the context of the
short-correlated flow.
It is worth noting that the above definition 
provides an operational method to measure~$t_{\mathrm{rel}}$ that can be
implemented in experiments. 

In summary, we have shown that the 
equilibrium configuration of a polymer  in a flow,
as well as the time a deformed molecule 
takes on average to recover that configuration, depend sensitively
on the properties of the flow.
In the vicinity of the coil--stretch transition 
the characteristic relaxation time
is much longer than the Zimm time,
both in elongational and random flows. 
In other words, the effective~\textit{Wi} differs considerably
from the ``bare'' one~\cite{note_2}. This effect is strongly amplified when the 
drag coefficient depends on  the conformation of the polymer,
and may play an important role in
drag-reducing turbulent flows,
where the strain rate often fluctuates around values typical of the 
coil--stretch transition~\cite{00SW}.
Our conclusions thus suggest  
that the conformation-dependent drag should be included as 
a basic ingredient of continuum models of polymer solutions,
calling for further theoretical, experimental and numerical 
study.

\acknowledgments
The authors gratefully acknowledge inspiring discussions with M. Chertkov, 
S.~Gerashchenko, M.~Martins Afonso  and V.~Steinberg. 
This work has been partially supported by the EU (contract
HPRN-CT-2002-00300).


\begin{thebibliography}{29}

\bibitem{chu}
S.~Chu, Phil. Trans. R. Soc. Lond. A \textbf{361}, 689 (2003).

\bibitem{larson}
R.~G.~Larson, J. Rheol. \textbf{49}, 1 (2005).

\bibitem{shaqfeh}
E.~S.~G.~Shaqfeh, J. Non-Newton. Fluid Mech. \textbf{130}, 1 (2005).

\bibitem{95GB}
A.~Gyr and H.~W.~Bewersdorff, {\itshape Drag reduction of Turbulent Flows
by Additives} (Kluwer Academic, Dordrecht, Boston, 1995)

\bibitem{05GCS}
S.~Gerashchenko, C.~Chevallard, and V.~Steinberg, Europhys. Lett. \textbf{71}, 221 (2005).

\bibitem{53R}
P.~E.~Rouse, Jr., J.~Chem.~Phys. \textbf{21}, 1272 (1953).

\bibitem{zimm}
B.~H.~Zimm, J. Chem. Phys. \textbf{24}, 269 (1956).

\bibitem{97QBC}
S.~R.~Quake, H.~Babcock, and S.~Chu, Nature \textbf{388}, 151 (1997).

\bibitem{94PQSC}
T.~T.~Perkins {\itshape et al.}, Science \textbf{264}, 822 (1994).

\bibitem{96MCVCC}
S.~Manneville {\itshape et al.}, Europhys. Lett. \textbf{36}, 413 (1996).

\bibitem{95MB}
Y.~Marciano and F.~Brochard-Wyart, Macromolecules \textbf{28}, 985 (1995).

\bibitem{02RZ}
R.~Rzehak and W.~Zimmermann, Europhys. Lett. \textbf{59}, 779 (2002).

\bibitem{74DG}
P.~G.~de~Gennes, J. Chem. Phys. \textbf{60}, 5030 (1974).

\bibitem{05CMV}
A.~Celani, S.~Musacchio, and D.~Vincenzi, J. Stat. Phys. \textbf{118}, 531 (2005).

\bibitem{05MV}
M.~Martins~Afonso and D.~Vincenzi, J. Fluid Mech. \textbf{540}, 99 (2005).

\bibitem{05VB}
D.~Vincenzi and E. Bodenschatz, J. Phys. A: Math. Gen. \textbf{39}, 10691 
(2006) 

\bibitem{TKCW}
V.~Tirtaatmadja, G.~H.~McKinley, and J.~J.~Cooper-White, Phys. Fluids
\textbf{18}, 043101 (2006).

\bibitem{03Schetal}
C.~M.~Schroeder {\it et al.}, Science \textbf{301}, 1515 (2003).

\bibitem{04SSC}
C.~M.~Schroeder, E.~S.~G.~Shaqfeh, and S.~Chu, Macromolecules \textbf{37},
9242 (2004). 

\bibitem{05HL}
C.-C.~Hsieh and R.~G.~Larson, J. Rheol. \textbf{49}, 1081
(2005).

\bibitem{Bird}
 R. B. Bird \textit{et al.},
{\itshape Dynamics of Polymeric Liquids}, vol. 2, 2nd edition 
(Wiley, New York, 1987).


\bibitem{99HQ}
J.~W.~Hatfield and S.~R.~Quake, Phys. Rev. Lett. \textbf{82}, 3548 (1999).

\bibitem{74H}
E.~J.~Hinch, in {\itshape Proceedings of 
Colloques Internationaux du CNRS No. 233}
(CNRS Editions, Paris, 1974), p.~241.

\bibitem{75T}
R. I. Tanner, Trans. Soc. Rheol. \textbf{19}, 557 (1975)

\bibitem{97L}
R.~G.~Larson {\it et al.}, Phys. Rev. E \textbf{55}, 1794 (1997).

\bibitem{Doyle}
P. S. Doyle  {\itshape et al.},
J. Non-Newton. Fluid Mech. \textbf{76}, 79 (1998).

\bibitem{note}
Our definition of~\textit{Wi} follows the literature on elongational
flows and therefore is half the one commonly adopted in the context of random flows.

\bibitem{MF}
P.~M. Morse and H.~Feshbach, {\it Methods of Theoretical Physics}
(McGraw--Hill, New York, 1953).

\bibitem{00GS}
A.~Groisman and V.~Steinberg, Nature \textbf{405}, 53 (2000);
New J. Phys. \textbf{6}, 29 (2004); 
T.~Burghelea, E.~Segre and V.~Steinberg, 
Phys. Fluids \textbf{17}, 103101 (2005).

\bibitem{Chertkov00}
M.~Chertkov, Phys. Rev. Lett. \textbf{84}, 4761 (2000);
E.~Bal\-kovsky, A.~Fouxon, and V.~Lebedev, Phys. Rev. Lett. \textbf{84}, 4765 
(2000);
J.-L. Thiffeault, Phys. Lett. A \textbf{308}, 445 (2003);
M.~Chertkov {\it et al.}, J. Fluid Mech. \textbf{531}, 251 (2005);
A.~Puliafito and K.~Turitsyn, Physica D \textbf{211}, 9 (2005);
J.~Davoudi and J. Schumacher, Phys. Fluids \textbf{18}, 025103 (2006).

\bibitem{68K}
R.~H.~Kraichnan, Phys. Fluids \textbf{11}, 945 (1968).

\bibitem{Ito}
The integration algorithm is based on the It\^o stochastic differential equation 
equivalent to Eq.~\eqref{eq:dumbbell}, $d\bm R=
\{\bm\kappa(t)\cdot\bm{R}-f(R){\bm R}/[2\tau\nu(R)]-R_0^2\nu'(R)\bm R/[2\tau\nu^2(R)R]\}dt+
\sqrt{R_0^2/[\tau \nu(R)]}\,d\bm W(t)$~\cite{Oettinger}; $\bm W(t)$ is the Brownian motion.

\bibitem{Oettinger}
H. C. \"Ottinger, \textit{Stochastic Processes in Polymeric Fluids}
(Springer--Verlag, Berlin, Heidelberg, 1996). 

\bibitem{brunk}
B.~K.~Brunk, D.~L.~Koch, and L.~W.~Lion, Phys. Fluids \textbf{9}, 2670 (1997);
J. Fluid Mech. \textbf{364}, 81 (1998).

\bibitem{05CPT}
A.~Celani, A.~Puliafito, and K.~Turitsyn, Europhys. Lett. \textbf{70}, 
464 (2005).

\bibitem{note_2}
This fact is related to the overestimation of the drag reducing~\textit{Wi}
encountered in numerical simulations~\cite{04D}.

\bibitem{04D}
Y.~Dubief et al., J. Fluid Mech. \textbf{514}, 271 (2004).

\bibitem{00SW}
K.~R.~Sreenivasan and C.~M.~White, J. Fluid Mech. \textbf{409}, 149 (2000);
E.~Balkovsky, A.~Fouxon, and V.~Lebedev, Phys. Rev. E \textbf{64}, 056301 
(2001); 
G. Boffetta, A.~Celani, and S. Musacchio, Phys. Rev. Lett. \textbf{91}, 
034501 (2003); V.~S.~L'vov 
{\it et al.}, {\it ibid.} \textbf{92}, 244503 (2004).


\end{thebibliography}
\end{document}